\documentclass[final,5p,times,twocolumn,sort&compress]{elsarticle}

\usepackage{graphicx}

\usepackage{amssymb}
\usepackage{amsthm}

\newcommand{\Wcm}{\,{\rm W \, cm^{-2}}}
\newcommand{\mum}{\,{\mu \rm m}}
\newcommand{\fes}{\,{\rm fs}}

\newcommand{\beq}{\begin{equation}}
\newcommand{\eeq}{\end{equation}}
\newcommand{\beqa}{\begin{eqnarray}}
\newcommand{\eeqa}{\end{eqnarray}}
\newcommand{\seta}{\setlength\arraycolsep{1pt}}
\newcommand{\bm}{\mathbf}

\journal{Nuclear Instruments and Methods Section A}

\begin{document}

\begin{frontmatter}



\title{Radiation Reaction Effects on Electron Nonlinear Dynamics \\
and Ion Acceleration in Laser-solid Interaction}


\author[unipi]{M. Tamburini\corref{cor}}
\ead{tamburini@df.unipi.it}

\author[unipi]{F. Pegoraro}

\author[MPI]{A. Di Piazza}

\author[MPI]{C. H. Keitel}

\author[ICT]{T. V. Liseykina}

\author[unipi,INO]{A. Macchi}

\cortext[cor]{Corresponding author}

\address[unipi]{Dipartimento di Fisica ``E. Fermi'', Universit\`a di Pisa, Largo Bruno Pontecorvo 3, I-56127 Pisa, Italy}
\address[MPI]{Max-Planck-Institut f\"ur Kernphysik, Saupfercheckweg 1, D-69117 Heidelberg, Germany}
\address[ICT]{Institute of Computer Technologies, SD-RAS, Novosibirsk, Russia and
Institute of Physics, University of Rostock, Germany}
\address[INO]{Istituto Nazionale di Ottica, CNR, research unit ``A. Gozzini'', Pisa, Italy}

\begin{abstract}
Radiation Reaction (RR) effects in the interaction of an ultra-intense
laser pulse with a thin plasma foil are investigated analytically and
by two-dimensional (2D3P) Particle-In-Cell (PIC) simulations.
It is found that the radiation reaction force leads to a significant
electron cooling and to an increased spatial bunching of both electrons and ions.
A fully relativistic kinetic equation including RR effects is discussed and
it is shown that RR leads to a contraction of the available phase space volume.
The results of our PIC simulations are in qualitative agreement
with the predictions of the kinetic theory.
\end{abstract}


\end{frontmatter}



\section{Introduction} \label{Intro}

Current laser systems may deliver intensities up to $10^{22} \Wcm$
\cite{Yanovsky} and intensities up to $10^{26} \Wcm$ are expected
at the Extreme Light Infrastructure (ELI).
In such ultrahigh-intensity regime and for  typical laser wavelength 
$\lambda \sim 0.8 \mum$ the motion of electrons in the laser field is
ultra-relativistic and Radiation Reaction (RR) effects may become important.
The RR force describes the back-action of the radiation emitted by an 
accelerated electron on the electron itself and accounts for
the loss of the electron energy and momentum due to the emission of such radiation. 
Apart from the need of including RR effects in the dynamics of 
laser-plasma interactions in the ultra-relativistic regime, the latter also 
offers for the first time the opportunity to detect
RR effects experimentally \cite{Keitel, DiPiazza1}.

In this paper we present an approach to a kinetic description of 
laser-plasma interactions where RR effects are included via the 
Landau-Lifshitz (LL) force \cite{LL}. Some properties of the kinetic equation with RR
are discussed and in particular it is proved that the RR force
leads to a \emph{contraction} of the phase space volume.
Then, PIC simulations are used to study RR effects on the
acceleration of a thin plasma foil in the regime of Radiation Pressure dominance \cite{Esirkepov}.
Numerical simulations \cite{Esirkepov} suggested that Radiation Pressure 
Acceleration (RPA) becomes the dominant mechanism of ion acceleration at 
intensities exceeding $10^{23} \Wcm$.
Such RPA regime is attractive because of the foreseen high efficiency,
the quasi-monoenergetic features expected in the ion energy spectrum and the possibility to achieve a 
potentially ``unlimited'' acceleration \cite{Bulanov}.
Previous Particle-In-Cell (PIC) simulations \cite{Zhidkov}
showed signatures of RR effects at intensities exceeding
$5 \times 10^{22} \Wcm$ and increasing nonlinearly with the laser intensity.
More recent simulations studies of RPA both for thick targets 
\cite{Naumova, Schlegel} and ultrathin targets \cite{Pukhov} suggested that
the inclusion of the RR force cools the electrons and may improve
the quality of the ion spectrum. 

Our approach to the inclusion of RR effects in a PIC code has been discussed
in detail in Ref.\cite{Tamburini} where one-dimensional (1D) simulations 
of RPA have been also reported. In the present paper we report both additional
1D simulations and first two-dimensional (2D) simulations using
parameters similar to those of Ref.\cite{Pegoraro} where, in particular,
the impact of a Rayleigh-Taylor-like instability on
a thin foil acceleration was studied.

In classical electrodynamics, the effect of RR can be
included by means of the LL force \cite{LL}
{\seta
\beqa
\bm{F}_R =  - \left( \frac{4\pi}{3}\frac{r_e}{\lambda} \right)  
& \cdot & \left\{ 
\gamma \Big[ \Big( \frac{\partial}{\partial t}
+ \mathbf{v} \cdot \nabla \Big) \mathbf{E} + \mathbf{v} \times \Big( \frac{\partial}{\partial t}
+ \mathbf{v} \cdot \nabla \Big) \mathbf{B} \Big] \nonumber \right.\\
& - &
\Big[ \Big( \mathbf{E} + \mathbf{v} \times \mathbf{B} \Big) \times \mathbf{B} +
\Big( \mathbf{v} \cdot \mathbf{E} \Big) \mathbf{E} \Big]  \nonumber \\
& + &
\left.\gamma^2\Big[ \Big( \mathbf{E} + \mathbf{v} \times \mathbf{B} \Big)^2
- \Big( \mathbf{v} \cdot \mathbf{E} \Big)^2  \Big] \mathbf{v} \right\}, \label{3dLL}
\eeqa}
where $\mathbf{v}$ is the electron velocity, $\gamma$ is the relativistic factor,
$r_e \equiv e^2 / m c^2 \approx 2.8 \times 10^{-9} \mum$
is the classical electron radius, $\lambda=2\pi c/\omega$ is the 
laser wavelength and we use dimensionless quantities as in the PIC code:
time, space and momentum are normalized in units of $\omega^{-1}$,
$c \omega^{-1}$ and $m c$, respectively.
Consequently, EM fields are normalized in units of $m \omega c/|e|$
and densities in units of the critical density $n_c=m\omega^2/4\pi e^2$.

The LL approach holds in the classical framework and quantum effects are neglected.
As pointed out in \cite{Tamburini}, the first term of the LL force Eq.(\ref{3dLL})
i.e. the one containing the \emph{derivatives} of the electric and magnetic fields,
should be neglected because its effect is smaller than quantum effects
such as the spin force.
However, in Sec.(\ref{Kin}) we show the effect
of each term of the LL force Eq.(\ref{3dLL})
on the rate of change of the phase space volume.

\section{The kinetic equation with Radiation Reaction} \label{Kin}

In this section, a fully relativistic kinetic equation
that includes the RR effects is discussed.
We show a few basic properties of the kinetic equation
pointing out the peculiarities of the RR force whose
main new feature is that it \emph{does not} conserve the phase-space volume.

Generalized kinetic equations for non-conservative forces
and in particular for the RR force have been known since late sixties
for the Lorentz-Abraham-Dirac (LAD) equation \cite{Hakim1, Hakim2} and
late seventies for the LL equation \cite{Kuzmenkov}.
Recently, the generalized kinetic equation with the LL force included
has been used to study the RR effects on thermal electrons
in a magnetically confined plasma \cite{Hazeltine1} and to
develop a set of closed fluid equations with RR \cite{Hazeltine2,Hazeltine3,Berezhiani1}.
In this paper, we give the kinetic equation in a non-manifestly covariant form,
see \cite{Kuzmenkov, Hazeltine1} for the kinetic equation
in a manifestly Lorentz-covariant form.

The relativistic distribution function $f = f(\bm{q}, \bm{p}, t)$
evolves according to the collisionless transport equation
\beq  \label{kinetic}
\frac{\partial f}{\partial t} + \nabla_{\bm{q}} \cdot (f \, \bm{v}) + \nabla_{\bm{p}} \cdot (f \, \bm{F}) = 0 \, ,
\eeq
where $\bm{q}$ are the spatial coordinates, $\bm{v} = \bm{p} / \gamma$
is the three-dimensional velocity, $\gamma = \sqrt{1 + \bm{p}^{2}}$
is the relativistic factor and $\bm{F} = \bm{F}_L + \bm{F}_R$
is the mean force due to external and collective fields
($\bm{F}_L \equiv - (\bm{E} + \bm{v} \times \bm{B})$ is the Lorentz force
and $\bm{F}_R$ is given in Eq.(\ref{3dLL})).
Physically, Eq.(\ref{kinetic}) implies the conservation
of the number of particles.

The new key feature compared to the usual Vlasov equation
is that for the RR force $\bm{F}_{R}$ we have $\bm{\nabla}_{\bm{p}} \cdot \bm{F}_{R} \neq 0$.
Using Lagrangian coordinates $\bm{q}(t), \, \bm{p}(t)$,
Eq.(\ref{kinetic}) can be recast in the equivalent form
\beq \label{kinetic2}
\frac{d \ln f}{d t} = - \nabla_{\bm{p}} \cdot \bm{F} \, .
\eeq
According to Eq.(\ref{kinetic2}),
$\bm{\nabla}_{\bm{p}} \cdot \bm{F}$
provides the percentage of variation of the distribution function $f$
within the characteristic time scale $\omega^{-1}$.
Integrating Eq.(\ref{kinetic2}) along its characteristics,
we find that the distribution function $f$ remains positive as required.

Introducing the entropy density in the phase space
$s(\bm{q}, \bm{p}, t) = - f(\bm{q}, \bm{p}, t) \ln f(\bm{q}, \bm{p}, t)$,
from Eq.(\ref{kinetic}) we get the equation for the evolution
of the entropy density
\beq  \label{entropy}
\frac{\partial s}{\partial t} + \bm{\nabla}_{\bm{q}} \cdot (s \, \bm{v})
+ \nabla_{\bm{p}} \cdot (s \, \bm{F}) = f \, \nabla_{\bm{p}} \cdot \bm{F} \, .
\eeq
Integrating Eq.(\ref{entropy}) in the phase space,
we get the rate of variation of the total entropy $S(t)$
\beq \label{totalS}
\frac{d S(t)}{d t} = \int{d^3 q \, d^3 p \; f \, \nabla_{\bm{p}} \cdot \bm{F}} \, .
\eeq
The Lorentz force $\bm{F}_L \equiv - (\bm{E} + \bm{v} \times \bm{B})$
gives $\bm{\nabla}_{\bm{p}} \cdot \bm{F}_L = 0$ identically
thus $\bm{\nabla}_{\bm{p}} \cdot \bm{F} = \bm{\nabla}_{\bm{p}} \cdot \bm{F}_R$.
Moreover, the distribution function $f(\bm{q}, \bm{p}, t)$ is always non-negative
$f \geq 0$ thus the sign of $d S / d t$ is given by $\nabla_{\bm{p}} \cdot \bm{F}_R$ solely.

From the LL force Eq.(\ref{3dLL}) we get \cite{Tamburini2}
{\seta
\beqa
\bm{\nabla}_{\bm{p}} \cdot \bm{F}_R \; & = & \;
- \left( \frac{4 \pi}{3} \frac{r_e}{\lambda} \right)
\Bigg\{ \Big[ \bm{\nabla}_{\bm{q}} \cdot \bm{E}
- \bm{v} \cdot \Big( \bm{\nabla}_{\bm{q}} \times \bm{B} - \frac{\partial \bm{E}}{\partial t} \Big) \Big] +
\nonumber \\
2 \Big[ \frac{\bm{E}^2 + \bm{B}^2}{\gamma} \Big] 
& + & 4 \gamma \Big[ \Big(\bm{v} \times \bm{E} \Big)^2 + \Big( \bm{v} \times \bm{B} \Big)^2
- 2 \bm{v} \cdot \Big(\bm{E} \times \bm{B} \Big) \Big] \Bigg\} . \label{phsp}
\eeqa}
In a plasma, the kinetic equation is coupled
with the Maxwell equations for the self-consistent fields
{\seta
\beqa
\bm{\nabla}_{\bm{q}} \cdot \bm{E} & = & \frac{\rho}{\rho_c}
= \frac{1}{n_c} \sum_{j=e,i} Z_j \int{d^3 p \: f_j (\bm{q}, \bm{p}, t)}
\\
\bm{\nabla}_{\bm{q}} \times \bm{B} - \frac{\partial \bm{E}}{\partial t} & = &  \frac{\bm{j}}{j_c}
= \frac{1}{n_c c} \sum_{j=e,i} Z_j \int{d^3 p \: \bm{v} f_j (\bm{q}, \bm{p}, t)},
\eeqa}
where $\rho_c \equiv |e| n_c$, $j_c \equiv |e| n_c c$,
$\int{d^3 q \, d^3 p \: f_j (\bm{q}, \bm{p}, t)} = N_j$ is the total
number of particles for each species ($j=e$ electrons, $j=i$ ions)
and $Z_j$ is the charge of the particle species in units of
$|e|$ (for electrons $Z_e = -1$).
For a plasma, Eq.(\ref{phsp}) can be recast as
{\seta
\beqa
\bm{\nabla}_{\bm{p}} \cdot \bm{F}_R & = &
- \left( \frac{4 \pi}{3} \frac{r_e}{\lambda} \right)
\Bigg\{ \Big[ \frac{\rho}{\rho_c} - \bm{v} \cdot \frac{\bm{j}}{j_c} \Big]
+ 2 \Big[ \frac{\bm{E}^2 + \bm{B}^2}{\gamma} \Big] \nonumber \\
 & + & 4 \gamma \Big[ \Big(\bm{v} \times \bm{E} \Big)^2 + \Big( \bm{v} \times \bm{B} \Big)^2
- 2 \bm{v} \cdot \Big(\bm{E} \times \bm{B} \Big) \Big] \Bigg\} . \label{phspcontraction}
\eeqa}
The terms of Eq.(\ref{phspcontraction}) proportional
to the charge density $\rho$ and to the current density $\bm{j}$
come from the first term of the LL force Eq.(\ref{3dLL})
i.e. the term containing the derivatives of the fields.
In general, these terms can give either a positive
or negative contribution to $\bm{\nabla}_{\bm{p}} \cdot \bm{F}_R$.
%
%
%
The second term of Eq.(\ref{phspcontraction}) i.e. the term
proportional to $(\bm{E}^2 + \bm{B}^2)$ has always
a negative sign, its effect decreases with increasing
electron energy and it is typically negligible.
The third term of Eq.(\ref{phspcontraction}) comes from the
strongly anisotropic ``friction'' term of the LL force
i.e. the term proportional to $\gamma^2$ in Eq.(\ref{3dLL})
(see \cite{Tamburini} for a detailed discussion of this term)
and dominates in the ultra-relativistic limit $\gamma \gg 1$.

It is possible to prove \cite{Tamburini2} the following statement:
for any $\bm{v}$ such that $|\bm{v}| \in [0,1[$ then
\beq \label{proof}
\left[ \left(\bm{v} \times \bm{E} \right)^2 + \left(\bm{v} \times \bm{B} \right)^2
-2 \bm{v} \cdot \left( \bm{E} \times \bm{B} \right) \right]
+ \left[ \frac{\bm{E}^2 + \bm{B}^2}{2 \gamma^2}\right] \geq 0 \, ,
\eeq
therefore according with Eqs.(\ref{totalS}, \ref{phspcontraction}),
the terms of the LL force Eq.(\ref{3dLL}) that \emph{do not} depend on
the derivatives of the fields always lead to a
\emph{contraction} of the available phase space volume.
In a few special cases, the effect of the terms
of the LL force Eq.(\ref{3dLL}) that depend on the derivatives
of the fields (i.e. the terms proportional to $\rho$ and $\bm{j}$ in Eq.(\ref{phspcontraction}))
might lead to an expansion of the phase space volume.
Anyway, their effect should be negligible
compared to quantum effects as discussed in \cite{Tamburini}.

We show explicitly the contraction of the phase space in the special case
of a small bunch of electrons interacting with
a plane wave where collective fields are
assumed to be negligible compared with the plane wave fields.
Assuming an initial distribution
$f = g(\bm{q}) \, \delta^3 (\bm{p} - \bm{p}_0)$,
from Eqs.(\ref{totalS}, \ref{phspcontraction}) we have
{\seta
\beqa
\frac{d S(t)}{d t} & = & - \left( \frac{4 \pi}{3} \frac{r_e}{\lambda} \right)
\int d^3 q \; g(\bm{q}) \left\{ 2 \left[ \frac{\bm{E}^2 + \bm{B}^2}{\gamma(\bm{p}_0)} \right]
+ 4 \gamma(\bm{p}_0) \cdot \right. \nonumber \\
& \cdot & \left. \left[ \left( \bm{v}_0 \times \bm{E} \right)^2 + \left( \bm{v}_0 \times \bm{B} \right)^2
- 2 \bm{v}_0 \cdot \left( \bm{E} \times \bm{B} \right) \right] \right\} ,
\label{plane}
\eeqa}
where $\bm{v}_0 = \bm{p}_0 / \gamma(\bm{p}_0) $.
If the electron bunch counter-propagates with the plane wave
($[\bm{v}_0 \cdot \left( \bm{E} \times \bm{B} \right)] < 0$)
or propagates in the transverse direction
($[\bm{v}_0 \cdot \left( \bm{E} \times \bm{B} \right)] = 0$),
from Eq.(\ref{plane}) it is clear that RR leads to a
contraction of the phase space.
In particular, in the case of counter-propagation
(using $|\bm{E}| = |\bm{B}|$, $\bm{E} \cdot \bm{B}=0$)
we have $\bm{\nabla}_{\bm{p}} \cdot \bm{F}_R = - \left( 4 \pi r_e / 3 \lambda \right)
4 \bm{E}^2 \left[2 \gamma(\bm{p}_0) |\bm{v}_0| (1 + |\bm{v}_0|) + 1/\gamma(\bm{p}_0) \right]$.
On the other hand, if the bunch propagates in the same
direction of the plane wave ($\bm{v}_0$ parallel to $\bm{E} \times \bm{B}$),
then the contribution of the friction term (proportional to $\gamma$
in Eq.(\ref{phspcontraction})) becomes comparable with
the contribution of the second term (proportional to $(\bm{E}^2 + \bm{B}^2)$
in Eq.(\ref{phspcontraction}))
and we have $\bm{\nabla}_{\bm{p}} \cdot \bm{F}_R =
- \left( 4 \pi r_e / 3 \lambda \right)
\left[ 4 \bm{E}^2 / (1 + |\bm{v}_0|)^2 \gamma^3 (\bm{p}_0) \right]$
which still leads to a contraction of the phase space
but with a rate $\gamma^4$ smaller than the case of counter-propagation.
This reinforces the evidence of the strongly \emph{anisotropic} features
of the LL force Eq.(\ref{3dLL}) (see \cite{Tamburini} for further details).

The physical interpretation of the above properties is that
the RR force acts as a cooling mechanism for the system:
part of the energy and momentum are radiated away and
the spread in both momentum and coordinate space may be reduced.
This general prediction is confirmed by our PIC simulations
(see Sec.\ref{PIC}) where we found that RR effects
lead to both an increased bunching in space and to
a noticeable cooling of hot electrons.

Finally, it is worthwhile mentioning that Eq.(\ref{kinetic})
is more general than the Vlasov equation but the PIC approach
is still valid i.e. the PIC approach provides a solution for Eq.(\ref{kinetic})
and it not limited to the Vlasov equation \cite{Tamburini2}.

\section{PIC simulations} \label{PIC}

Suitable approximations to the LL force and our approach to its inclusion 
in a PIC code are described in Ref.\cite{Tamburini}. The numerical approach is
based on the widely used Boris particle pusher and it can be implemented in
codes of any dimensionality. Inclusion of RR effects via this method in PIC
simulations leads to only a $\sim 10\%$ increase in CPU time, which may be
essential to perform large-scale simulations with limited computing power.

\subsection{1D simulations}

We first report one-dimensional (1D3P) PIC simulations with laser and plasma 
parameters similar to Ref.\cite{Esirkepov}. Previous 1D simulations in this
regime have been reported in Ref.\cite{Tamburini} where a detailed comparison 
with other work is also made. In the present paper we review the basic 
observations in the 1D case and we include results at intensities higher than
those investigated in Ref.\cite{Tamburini}.

The target is a plasma foil of protons
with uniform initial density $n_0 = 100 n_c$ and thickness $\ell = 1 \lambda$
where $\lambda = 0.8 \mum$ is the laser wavelength
and $T = \lambda / c \approx 2.67 \fes$ is the laser period.
In these simulations, the laser pulse front reaches the edge of the plasma
foil at $t = 0$, the profile of the laser field amplitude has
a ``trapezoidal'' shape in time with one-cycle, $\sin^2$-function
rise and fall and a five cycles constant plateau.
We considered three intensities $I = 2.33 \times 10^{23} \Wcm$,
$I = 5.5 \times 10^{23} \Wcm$ and $I = 10^{24} \Wcm$ for both Circular (CP)
and Linear (LP) polarization of the laser pulse.

\begin{figure}[h!t]
\includegraphics[width=0.48\textwidth]{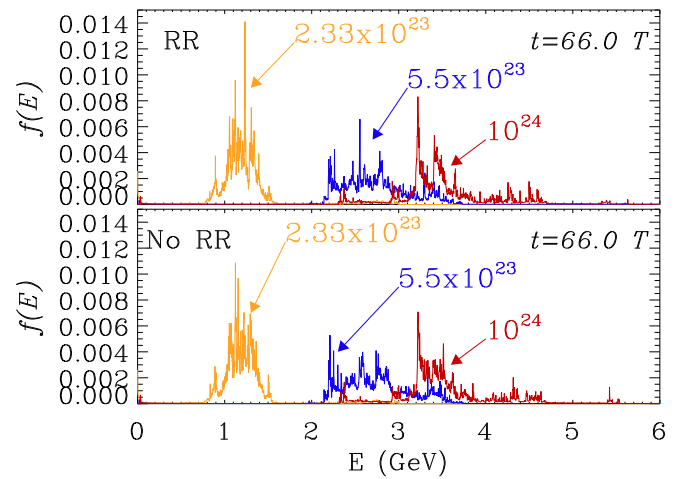}
\caption{\label{CPspectra} Ion energy spectra at $t = 66T$
with (top) and without (bottom) RR for Circular Polarization.
The laser intensity $I$ is $2.33 \times 10^{23} \Wcm$ (yellow),
$5.5 \times 10^{23} \Wcm$ (blue), $10^{24} \Wcm$ (red)
and the target thickness is $\ell = 1 \lambda$.}
\end{figure}

In the CP case, we found that RR effects on the ion spectrum
are negligible even at intensities of $I = 10^{24} \Wcm$ as shown in 
Fig.\ref{CPspectra}.
For CP, electrons pile up and the numerical density grows
exceeding thousand of times the critical density $n_c$.
The laser pulse does not penetrate deeply into the target
(i.e. the effective skin depth is a very small fraction of the foil thickness)
and electrons move in a field much weaker than the vacuum field.

In Ref.\cite{Esirkepov} it was expected that RR effects in the 
radiation-pressure dominated acceleration of the thin foil would have been 
weak because in this regime the whole foil becomes quickly relativistic,
hence in the foil frame the laser wavelength $\lambda'$ increases and the
typical strength of the RR parameter $\sim  r_e/\lambda$ [see Eq.(\ref{3dLL})] decreases.
The present case of acceleration with CP pulses appears to confirm this
picture. The weakness of RR effects may also be explained on the basis of the
LL equation for an electron moving into a plane wave \cite{DiPiazza}.
As electrons move in the forward direction coherently with the foil (while 
rotating in the transverse plane in the CP field) and the amplitude of the 
reflected wave is weak when the foil is strongly relativistic, the situation is 
similar to an electron co-propagating with the plane wave at a velocity close 
to $c$, for which the LL force almost vanishes \cite{Tamburini}.
The relativistic motion of the foil also prevents the onset of 
Self-Induced Transparency by increasing the optical thickness 
parameter $\zeta = \pi n_0 \ell / n_c \lambda$
in the foil frame (see \cite{Macchi} and references therein) .
For smaller target thickness, breakthrough of the laser pulse occurs and 
RR effects are greatly enhanced also for CP \cite{Tamburini}.

It may be worth noticing that, at the highest intensity considered  
$I = 10^{24} \Wcm$, in principle one would expect the classical approach to 
RR to break down due to the onset of quantum electrodynamics (QED) effects,
as discussed in Ref.\cite{Tamburini}.
However, it can be shown by a direct analysis of the simulation data
that the threshold condition for QED effect is not violated in the CP case.

\begin{figure}[h!t]
\includegraphics[width=0.48\textwidth]{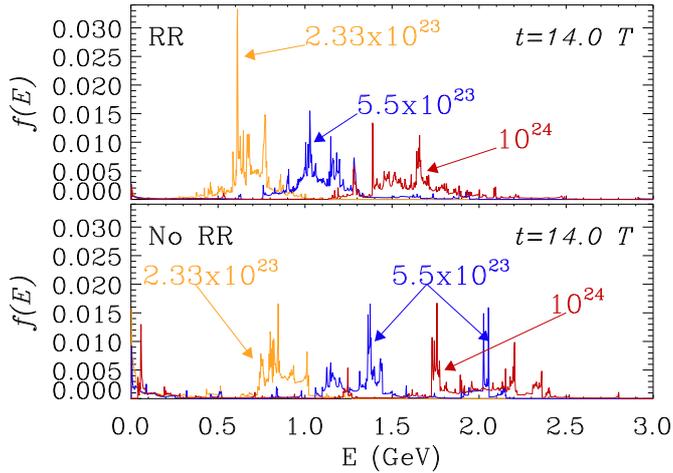}
\caption{\label{LPspectra} Ion energy spectra at $t = 14T$
with (top) and without (bottom) RR for Linear Polarization.
The laser intensity $I$ is $2.33 \times 10^{23} \Wcm$ (yellow),
$5.5 \times 10^{23} \Wcm$ (blue), $10^{24} \Wcm$ (red)
and the target thickness is $\ell = 1 \lambda$.}
\end{figure}

For linear polarization (LP), differently from the CP case,
we found that RR effects are important leading to a reduction
of the maximum achievable ion energy and to some narrowing of
the width of the ion spectrum as shown in Fig.\ref{LPspectra}.
This different dynamics for LP is correlated with the strong longitudinal
oscillatory motion driven by the oscillating component
of the $\bm{j} \times \bm{B}$ force which is suppressed
in the CP case. This allows a deeper penetration
of the laser pulse into the foil with a significant
fraction of electrons on the front surface moving
in a strong electromagnetic field of the same order
of vacuum fields \cite{Tamburini}.
The relative reduction in the ion energy when RR is included
is close to the percentage of the laser pulse energy which
is lost as high-energy radiation escaping from the plasma.

The results for LP (Fig.\ref{LPspectra}) are shown for
the same intensity values of the CP case (Fig.\ref{CPspectra})
for a direct comparison.
However, at least for the highest intensity case,
the LP results must be taken with some caution as 
the condition for the validity of a classical approach
tends to be significantly violated.
In such regime, an analysis based on quantum RR effects
might be necessary \cite{DiPiazza2, Sokolov1}.

\subsection{2D simulations}

We report preliminary two-dimensional (2D3P) PIC simulations
with laser and plasma parameters similar to Ref.\cite{Pegoraro}.
To the best of our knowledge, this is the first paper reporting
results of two-dimensional PIC simulations with RR
effects included.

\begin{figure}[h!t]
\includegraphics[width=0.48\textwidth]{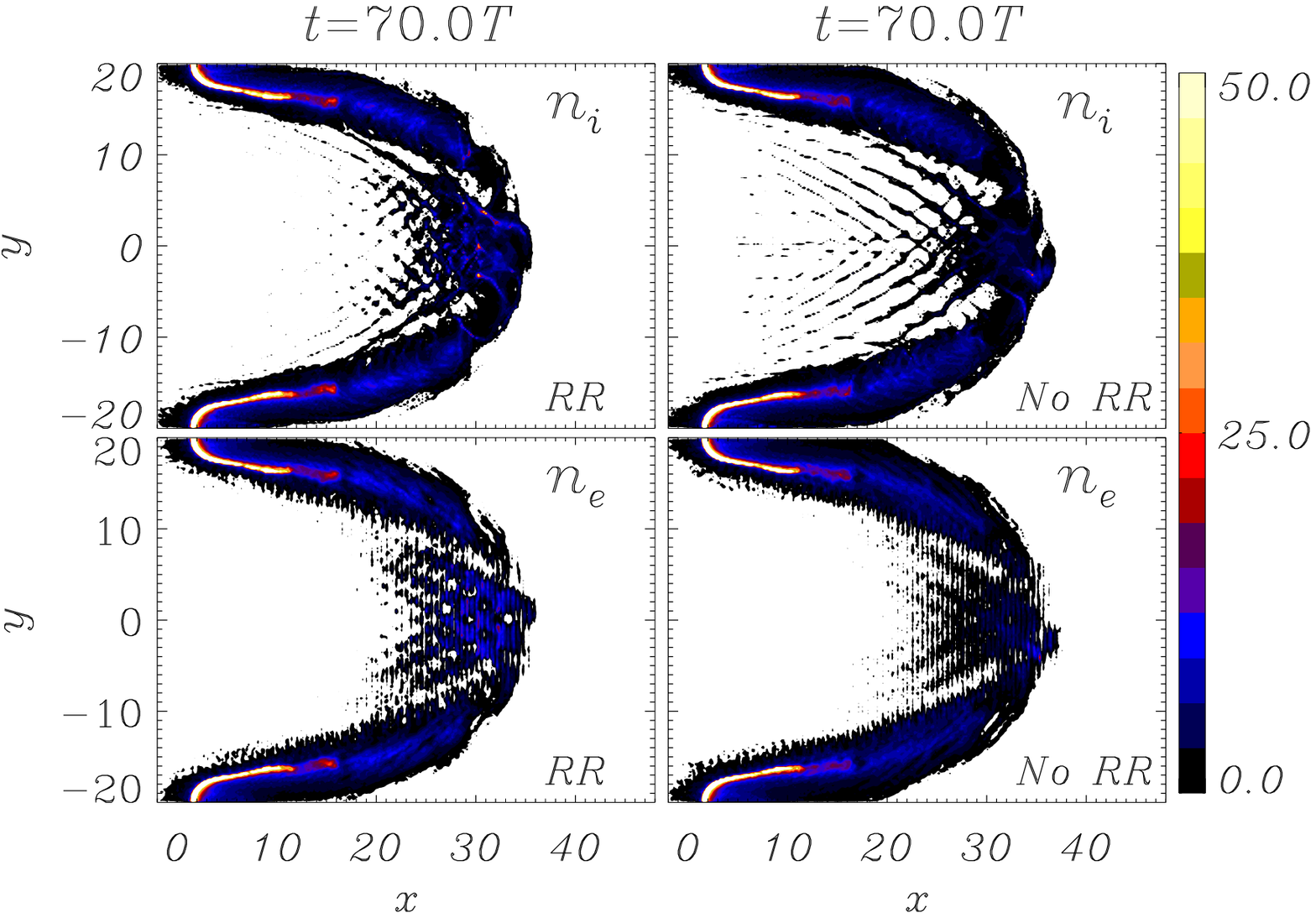}
\includegraphics[width=0.48\textwidth]{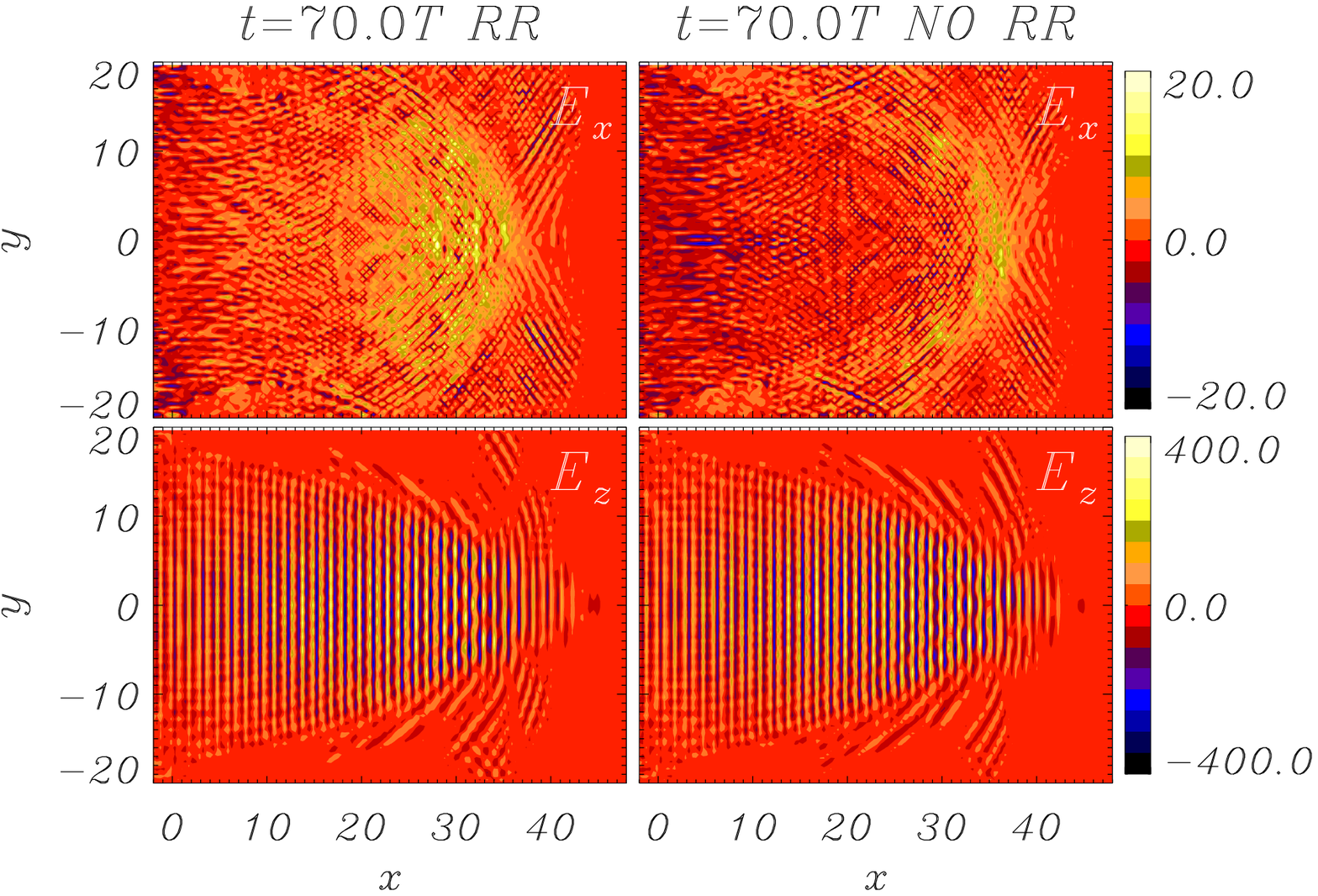}
\includegraphics[width=0.48\textwidth]{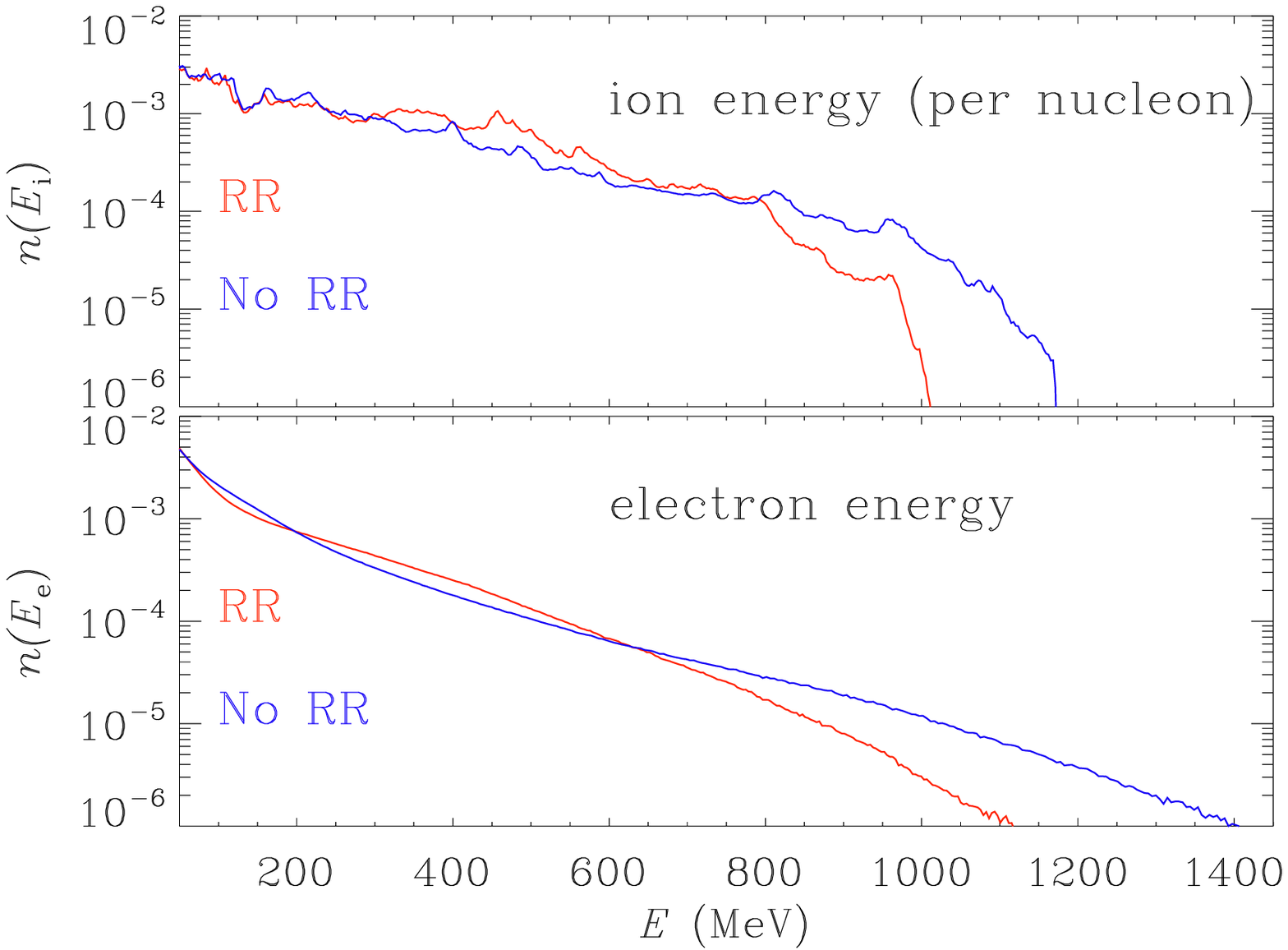}
\caption{\label{2Dsim} Plots of the 2D PIC simulations at $t=70T$.
The laser pulse is s-polarized with an intensity $I = 1.4 \times 10^{23} \Wcm$
and the target thickness is $\ell = 0.5 \lambda$.
From top to bottom, ion $n_i$ and electron $n_e$ density distributions
with (left column) and without (right column) RR, longitudinal $E_x$ (first row)
and transverse $E_z$ (second row) electric field, ion and electron spectrum
with (red) and without (blue) RR.}
\end{figure}

The target is a plasma slab of fully ionized deuterium ($Z/A=1/2$)
of width $40 \lambda$, density $n_0 = 169 n_c$
and thickness $\ell = 0.5 \lambda$.
The size of the computational box
is $95 \lambda \times 40 \lambda$ with a
spatial resolution $\Delta x = \Delta y = \lambda / 80$
and $625$ quasi-particles per cell corresponding
to a total of $8 \times 10^{7}$ quasi-particles.
The laser pulse is s-polarized with the electric field
along the $z$-axis. Its normalized amplitude is $a_0 = 320$
corresponding to an intensity $I = 1.4 \times 10^{23} \Wcm$
with a wavelength $\lambda = 1.0 \mum$ and period $T = \lambda / c \approx 3.3 \fes$.
The pulse has a Gaussian transverse profile
of width $20 \lambda$ FWHM and a $\sin^2$ longitudinal profile
of length $40 \lambda$ FWHM.
In these simulations, the front of the lase pulse
reaches the foil at $t=0$.

Comparing the results of our simulations
with and without RR (see Fig.\ref{2Dsim}, we report the results at $t = 70 T$)
it is apparent that RR leads to both an increased electron
and ion bunching and to a strong cooling of electrons.
These results are qualitatively consistent with our expectations
from the kinetic theory that we have discussed in Sec.\ref{Kin}
and in particular with the prediction of a contraction
of the electrons available phase space volume.

A qualitative understanding of these results
can be achieved recalling that the RR
force Eq.(\ref{3dLL}) is mainly a strongly anisotropic and non-linear
friction-like force that reaches its
maximum for electrons that counter-propagate
with the laser pulse \cite{Tamburini}.
The backward motion of electrons is thus impeded by RR,
more electrons and consequently ions are pushed
forward leading to an enhanced clumping
that improves the efficiency of the RPA mechanism.
In fact, the ion spectrum with RR shows a region
between about three hundred and six hundred MeV with a significant
increase in the number of ions compared to the case without RR (Fig.\ref{2Dsim}).
This picture is confirmed by both the enhancement of the longitudinal
electric field $E_x$ and by the formation of denser bunches
in the ion density compared to the case without RR (see Fig.\ref{2Dsim}).
However, for linear polarization,
hot electrons are always generated by the
oscillating component of the $\bm{j} \times \bm{B}$ force.
The generation of hot electrons provides a competing
acceleration mechanism to RPA and ultimately leads
to the generation of the fraction of ions with the highest energy.
The noticeable suppression of the $\bm{j} \times \bm{B}$
heating mechanism due to the RR force
therefore leads to a lower maximum cut-off energy
both in the electron and in the ion spectrum (see Fig.\ref{2Dsim}).

These preliminary results for two-dimensional simulations
with RR effects included suggest that,
in the LP case, the trends found
in one-dimensional simulations
hold qualitatively even for higher dimensions.
More detailed studies and quantitative comparisons
between one-dimensional and two-dimensional PIC
simulations are left for forthcoming publications.

\section{Conclusions} \label{Conclusions}

We summarize our results as follows.
Radiation Reaction effects on the electron dynamics
in the interaction of an ultra-intense laser pulse
with a thin plasma foil were studied analytically and
by one-dimensional and two-dimensional PIC simulations.
The details of the numerical implementation
of the RR force in our PIC code were described
in Ref.\cite{Tamburini}.

In one-dimensional simulations, we checked
RR effects for three different intensities:
$I = 2.33 \times 10^{23} \Wcm$,
$I = 5.5 \times 10^{23} \Wcm$ and $I = 10^{24} \Wcm$
comparing the results for Circular and Linear Polarization
of the laser pulse.
For CP, we found that RR effects are not relevant
even at intensity of $I = 10^{24} \Wcm$ whenever
the laser pulse does not break through the foil.
In contrast, for LP we found that RR effects are
important reducing the ion energy significantly.

In two-dimensional simulations, we found that
RR reduces the $\bm{j} \times \bm{B}$
heating mechanism leading to a lower maximum
cut-off energy both in the electron and in the ion spectrum.
Moreover, RR increases the spatial bunching
of both electrons and ions which are collected
into denser clumps compared to the case without RR.
This might lead to a somewhat beneficial effect
with a longer and more efficient radiation pressure
acceleration phase whose signature
would be an ion energy spectrum peaking
at an intermediate energy.

A generalized relativistic kinetic equation including
RR effects has been discussed and we have shown
that RR leads to a contraction of the available
phase space volume.
This prediction is in qualitative agreement
with the results of our PIC simulations
where we observed both an increased spatial bunching
and a significant electron cooling as discussed above.

\section*{Acknowledgments}

We acknowledge the CINECA award under the ISCRA initiative
(project ``TOFUSEX''), for the availability
of high performance computing resources and support.








\end{document}